\documentclass[lettersize,journal]{IEEEtran}
\usepackage{cite}
\usepackage{amsmath}
\usepackage{amssymb}
\usepackage{algorithmic}
\usepackage{array}
\usepackage[pdftex]{graphicx}
\usepackage{caption}
\usepackage[most]{tcolorbox}
\usepackage{soul}
\usepackage{adjustbox}
\DeclareGraphicsExtensions{.pdf}
\usepackage{subcaption}
\usepackage{multirow}
\usepackage{balance}
\usepackage{booktabs}
\usepackage{cleveref}

\ifCLASSOPTIONcompsoc
\usepackage[caption=false,font=normalsize,labelfont=sf,textfont=sf]{subfig}

\else
 \usepackage[caption=false,font=footnotesize]{subfig}

\usepackage{stfloats}
\usepackage{url}
\begin{document}

\title{ARMATA: Auto-Regressive Multi-Agent Task Assignment}

\author{
    Yazan Youssef, \IEEEmembership{Member, IEEE}, Aboelmagd Noureldin, \IEEEmembership{Senior Member, IEEE}, and Sidney Givigi, \IEEEmembership{Senior Member, IEEE}%
    \thanks{Y. Youssef is with the Department
    of Electrical and Computer Engineering, Queen's University, Kingston, ON, Canada (e-mail: yazan.youssef@queensu.ca).}%
    \thanks{A. Noureldin is with the Department of Electrical and Computer Engineering at the Royal Military College of Canada, and the Department of Electrical and Computer Engineering at Queen's University, both in Kingston, Canada (e-mail: aboelmagd.noureldin@rmc.ca).}%
    \thanks{S. Givigi is with the School of Computing, Queen's University, Kingston, ON, Canada (e-mail: \{sidney.givigi, paulo.araujo\}@queensu.ca).}%
}
\maketitle
\begin{abstract}
Coordinating multi-agent systems over spatially distributed areas requires solving a complex hierarchical problem: first distributing areas among agents (allocation) and subsequently determining the optimal visitation order (routing). Existing methods typically decouple these stages ignoring inter-stage dependencies or rely on decentralized heuristics that lack global context. In this work, we propose a centralized, fully end-to-end auto-regressive framework that jointly generates allocation decisions and routing sequences. The core contribution of our approach is a multi-stage decoding mechanism that unifies high-level allocation and low-level routing in a single autoregressive pass while maintaining a centralized global state. This enables the model to implicitly balance workload distribution with routing efficiency, avoiding local optima common in decentralized methods. Extensive experiments demonstrate that our method significantly outperforms diverse baselines, achieving up to a 20\% improvement in solution quality over industrial solvers such as Google OR-Tools, IBM CPLEX, and LKH-3, while reducing computation time from hours to seconds. 
\end{abstract}


\IEEEpeerreviewmaketitle

\section{Introduction}\label{sec:introduction}
The deployment of Multi-Agent Systems (MAS) has become a cornerstone of modern automation, driving efficiency in domains ranging from large-scale logistics and warehouse automation to surveillance and disaster response~\cite{2024_MAS_survey}. Central to these applications is the Multi-Agent Task Assignment and Routing problem, where a fleet of agents must service a set of spatially distributed areas~\cite{2023_TAandRouting}. This problem presents a hierarchical challenge, as it involves two stages: (i) task assignment, and (ii) path planning or routing. In other words, each agent needs first to get assigned some areas, and then it needs to determine the optimal visitation order. As the scale of the environment and the size of the fleet increase, the interdependence between the high-level assignment decisions and the low-level routing creates a combinatorial explosion, rendering the problem NP-hard~\cite{2024_CoordinatedScheduling}.

Traditionally, this problem has been addressed using exact mathematical programming solvers, such as those based on Branch-and-Cut algorithms (e.g., IBM CPLEX~\cite{IBM_CPLEX}), or meta-heuristic libraries (e.g., Google OR-Tools~\cite{ortools})~\cite{2023_traditional_solvers}. While exact solvers guarantee optimality, they suffer from exponential time complexity, often requiring hours or days to converge on large-scale instances, making them impractical for real-time applications~\cite{2024_exactMethods}. Conversely, heuristic approaches offer faster solutions but are prone to getting trapped in local optima~\cite{2021_mataHeuristics}. This is largely because they tend to decouple the problem (i.e., solving allocation and routing sequentially) thereby ignoring the coupled nature of the objective function~\cite{Li2025}.

Recent advancements in Deep Reinforcement Learning (DRL) and Neural Combinatorial Optimization (NCO) have offered promising alternatives, demonstrating the ability to learn heuristic policies directly from data~\cite{2025_DRL_with_NCO}. Existing learning-based frameworks exhibit various limitations when applied to the multi-stage map coverage problem. The majority of current literature focuses either on decentralized architectures, where agents act with limited local observations, or on ``flat'' encoding schemes that fail to capture the hierarchical dependencies between task assignment and path planning~\cite{2025_survey_MAPF}. Decentralized methods, while scalable, often lack the global context necessary to optimize the collective objective, leading to redundant coverage or inefficient workload distribution~\cite{2025_wiley_decentralized}. Furthermore, few existing frameworks address the specific complexity of multi-stage decision-making, where an action involves both assignment and sequencing, within a unified end-to-end differentiable pipeline.

To address these challenges, we propose a novel centralized, fully end-to-end autoregressive framework. In contrast to prior approaches that rely on decentralized policies or decoupled stages, our approach leverages a global view of the environment to jointly optimize allocation and routing. The proposed architecture autoregressively generates a solution by decoding the hierarchical structure of the problem in a single pass. By maintaining a centralized state representation, the model effectively propagates gradients across both decision stages, ensuring that assignment and routing are optimized jointly.

This work bridges the gap between high-fidelity combinatorial optimization and real-time decision-making. The contributions of this paper can be summarized as:
\begin{itemize}
    \item \textbf{Centralized Multi-Stage Autoregressive Framework}: We propose the first fully end-to-end centralized autoregressive framework for jointly solving the coupled area-assignment and routing problem. By eliminating intermediate heuristics and explicit problem decomposition, this approach paves the way for a new research direction for unified learning-based solutions. 

    \item \textbf{Multi-Stage Decoding Mechanism}: We introduce a novel decoding architecture tailored to the hierarchical structure of the problem, allowing the simultaneous generation of optimal agent-task distributions and visitation sequences.

    \item \textbf{High Solution Quality}: Extensive benchmarking shows that the proposed framework outperforms well-established industrial solvers, like Google OR-Tools and IBM CPLEX, achieving up to a $20\%$ improvement in solution quality.

    \item \textbf{Real-Time Efficiency}: The proposed method achieves a speedup of four orders of magnitude over traditional exact solvers. While such solvers may require up to $44$ hours to converge on complex instances, our framework generates high-quality solutions within seconds, enabling dynamic replanning and real-time applicability.
\end{itemize}

The paper is organized as follows. Section~\ref{sec:related_work} reviews recent advances in multi-agent task assignment and path planning. Section~\ref{sec:framework} presents the proposed ARMATA architecture in detail. The simulation setup is described in Section~\ref{sec:SimulationSetup}, and the obtained results are presented and discussed in Section~\ref{sec:results}. Finally, Section~\ref{sec:conclusion} concludes the paper by summarizing the main findings and outlining directions for future work.
\section{Related Work}\label{sec:related_work}
Multi-agent task assignment and routing problems arise in numerous real-world applications, including logistics, warehouse automation, and multi-robot coordination~\cite{choudhury2022dynamic}. This section reviews solution approaches ranging from classical exact and heuristic methods to modern learning-based techniques.

\subsection{Heuristics and Exact Methods}
Classical approaches to vehicle routing and task assignment problems can be broadly categorized into exact methods and heuristic algorithms. Exact methods, including mixed-integer linear programming (MILP), branch-and-bound, branch-and-cut, and branch-and-price algorithms, guarantee optimal solutions but suffer from exponential computational complexity~\cite{2002_overview}. Besides, exact methods face fundamental scalability limitations. The NP-hard nature of assignment and routing problems means that computation time grows exponentially with problem size, making exact approaches impractical for real-time applications or large-scale instances~\cite{2016_ExactWeakness}.

To address these challenges, researchers developed heuristic and metaheuristic algorithms which sacrifice optimality guarantees for computational efficiency~\cite{2017_heuristics_survey}. Construction heuristics, such as the savings algorithm and nearest neighbor insertion, provide fast initial solutions by iteratively building routes~\cite{1996_heuristic}. Local search methods, including 2-opt, 3-opt, and Or-opt, improve solutions through neighborhood exploration~\cite{2011_optHeuristic}. Metaheuristics, like greedy descent, simulated anealing, and tabu search, have demonstrated strong performance on benchmark instances~\cite{2024_metaheuristic}.

However, heuristic methods exhibit some critical weaknesses as they cannot derive insights from historical problem instances, treating each new instance independently without leveraging patterns from past experience~\cite{2020_heuristic_weakness_I}. Furthermore, metaheuristics often require substantial computation time to converge to high-quality solutions, limiting their applicability in time-sensitive scenarios such as real-time routing~\cite{2019_heuristic_longtime}. Finally, extending these methods to handle additional constraints or problem variants typically requires substantial algorithmic redesign~\cite{2016_ExactWeakness}. These limitations motivate the development of learning-based approaches capable of generalizing across problem instances while providing rapid inference.

\subsection{Neural Combinatorial Optimization}
Neural Combinatorial Optimization (NCO) has emerged as a robust alternative to classical operations research methods, offering the ability to learn transferable heuristics that support rapid, real-time inference~\cite{2021_NeuralCO}. The structural foundation of NCO lies in sequence-to-sequence learning, pioneered by Vinyals et al.~\cite{vinyals2015pointer} via Pointer Networks and later refined by Kool et al.~\cite{kool2018attention} using Transformer-based Attention Models. These architectures leverage multi-head self-attention to encode global graph structures and autoregressively decode solutions, achieving competitive performance with significantly reduced computational overhead.

To eliminate the dependency on labeled optimal solutions, Bello et al.~\cite{bello2016neural} introduced reinforcement learning (RL) frameworks that directly optimize objectives via policy gradients. Subsequent research has addressed stability and generalization; notably, Kwon et al.~\cite{kwon2020pomo} exploited solution symmetries to reduce variance. Concurrently, efforts to improve cross-distribution and cross-scale generalization have led to innovations in heavy decoder architectures~\cite{luo2023neural}, bisimulation quotienting~\cite{drakulic2023bq}, and prompt learning for zero-shot adaptation~\cite{liu2024prompt}. These advancements make NCO uniquely positioned to address the complexities of multi-agent task assignment and routing in heterogeneous, dynamic environments.

\subsection{Multi-Agent Task Assignment and Routing}
Multi-agent task assignment and routing extend single-vehicle formulations to complex scenarios requiring the coordination of heterogeneous agents to complete spatially distributed tasks~\cite{eksioglu2009vehicle}. A fundamental challenge in this domain is the inherent interdependence of assignment and routing: optimal task allocation is contingent upon the resulting routes, while optimal path planning depends heavily on the assignment strategy~\cite{2021_mataHeuristics}. In~\cite{2021_mataHeuristics}, the joint optimization of these decisions has been shown to yield significantly superior solutions compared to sequential, decoupled approaches.

To address this coupling, recent literature has largely coalesced around the Centralized Training with Decentralized Execution (CTDE) paradigm~\cite{lowe2017multi}. CTDE balances global coordination with operational autonomy by leveraging global state information during training while restricting agents to local observations during execution. Using this framework, Liu et al.~\cite{liu2019task} demonstrated that jointly optimizing assignment and path planning minimizes total travel distance and makespan, while ensuring collision-free schedules. Dai et al.~\cite{dai2025heterogeneous} extended these capabilities to heterogeneous fleets, introducing a constrained flash-forward mechanism to balance short-term cooperation with long-term scheduling dependencies.

While CTDE addresses coordination, scalability and robustness to communication failures are often managed through Graph Neural Networks (GNNs) and decentralized protocols~\cite{niu2021multi}. GNNs have proven effective for encoding relational structures in multi-agent systems, facilitating decentralized target localization~\cite{peng2024graph} and tracking~\cite{zhou2022graph} via limited-hop message passing. Advanced frameworks integrate GNNs with CTDE to enable dynamic and asynchronous task allocation without reliance on central coordination~\cite{2025magnnet}. Similarly, Gabler et al.~\cite{gabler2024decentralized} utilized best-response policies to enable asynchronous decision-making suitable for real-world deployments.

Finally, attention-based NCO architectures have also been tailored to handle the combinatorial complexity of multi-agent routing. Berto et al.~\cite{berto2024parallel} proposed Parallel AutoRegressive Combinatorial Optimization, which employs parallel decoding with priority-based conflict handling to achieve performance competitive with heuristics. In heterogeneous fleet settings, Xiang et al.~\cite{xiang2024centralized} applied centralized DRL to dynamic pickup-and-delivery problems, effectively managing the coordination of diverse vehicle types.

Despite recent advances, existing methodologies face intrinsic limitations that this work seeks to address. Decentralized methods offer scalability but operate under partial observability, often converging to local optima that sacrifice global solution quality~\cite{xiao2025asynchronous}. Conversely, centralized learning approaches frequently decouple task assignment and routing into sequential subproblems or rely on parallel decoding schemes~\cite{2025_end_to_end_learning}. Such designs often lead to assignment conflicts, requiring computationally costly post-processing or heuristics resolution. The framework proposed in~\cite{2025_end_to_end_learning} represents an early attempt to jointly address task assignment and path planning. However, despite adopting an end-to-end learning approach, the formulation did not constitute a strict end-to-end framework. Moreover, it exhibited a high degeneration rate (i.e., agents not getting assigned any tasks) and could not generalize to new unseen problem instances of different sizes. 

To bridge this gap, we propose a unified, centralized, autoregressive, fully end-to-end framework that constructs solutions via sequential decision-making under full observability. This architecture offers three distinct advantages. First, by generating decisions sequentially, the model inherently coordinates agent actions, naturally preventing assignment conflicts without requiring explicit masking or post-resolution mechanisms. Second, the centralized formulation enables true joint optimization of assignment and routing, eliminating the suboptimality of decomposed approaches. Finally, the attention-based multi-stage decoder captures complex agent-task compatibilities and allows for the seamless integration of dynamic task arrivals by extending the decision sequence as new demands appear, allowing anticipatory planning under uncertainty within a single, cohesive policy.

\section{Framework}\label{sec:framework}
In this section, we present a multi-agent task assignment framework for scenarios in which agents are assigned to geographic locations, such as search-and-rescue operations. As a representative example, we consider a mission wherein $M$ agents are tasked with scanning $N$ distinct areas. Although the discussion focuses on search and rescue, the proposed framework readily extends to other applications, including warehouse management and drone delivery. In all cases, the objective is to compute an assignment strategy that best satisfies mission goals. In the search-and-rescue case, the best assignment specifies both (i) which areas each agent should cover and (ii) the scanning order or route, with the goal of minimizing total travel distance while ensuring full coverage.

Within the proposed framework, task allocation and waypoint planning are formulated as a sequencing problem. At each time step $t$, the task-allocation policy is represented by a matrix $\pi_{TA} \in \mathbb{R}^{M \times N}$, where each element $ta_{mn}$ denotes the probability of assigning agent $m$ to area $n$. Rather than solving this policy in a single step, decisions are generated sequentially in a fully autoregressive manner: agents are processed in a fixed order, and a task is assigned to each agent before proceeding to the next. This contrasts with alternative approaches that rely on an initial clustering stage~\cite{2025_end_to_end_learning}, in which agents are first assigned subsets of areas, followed by a separate routing optimization. By eliminating clustering and performing task assignment end-to-end, the proposed framework offers a new perspective on multi-agent allocation. In particular, it is well-suited for dynamic or online settings, where decisions must be adapted to newly available information. Moreover, the autoregressive approach can reduce the likelihood of settling into sub-optimal solutions, as agents are not constrained by pre-defined clusterS and can maintain broader exploration horizons throughout the assignment process.
\subsection{Input}\label{subsec:input}
The framework takes as input a map encoding agent locations and geometric representations of the areas be visited and scanned. This map is represented by four matrices: the agents' locations matrix $Z \in \mathbb{R}^{m \times 2}$, the area adjacency matrix $A \in \mathbb{R}^{n\times n}$, the area feature matrix $X \in \mathbb{R}^{n\times f}$, and the area position matrix $P \in \mathbb{R}^{n\times 2}$, where $m,n$ and $f$ are the number of agents, areas, and features, respectively.

Matrix $Z$ contains the ($x,y$) coordinates of agent locations, while $A$ encodes connectivity among areas. The feature matrix $X$ encodes task-relevant area attributes; in our setting, it consists of the ($x,y$) coordinates of the areas' corners. Lastly, $P$ provides the ($x,y$) coordinates of the areas' centers.
\subsection{Encoder} \label{sec:Encoder}
The encoder in ARMATA consists of two components: Nodes' encoder and agents' encoder.

\subsubsection{Nodes encoder}\label{subsubsec:node_enc}
\begin{figure*}
    \centering
    \resizebox{\textwidth}{0.4\textwidth}{\includegraphics{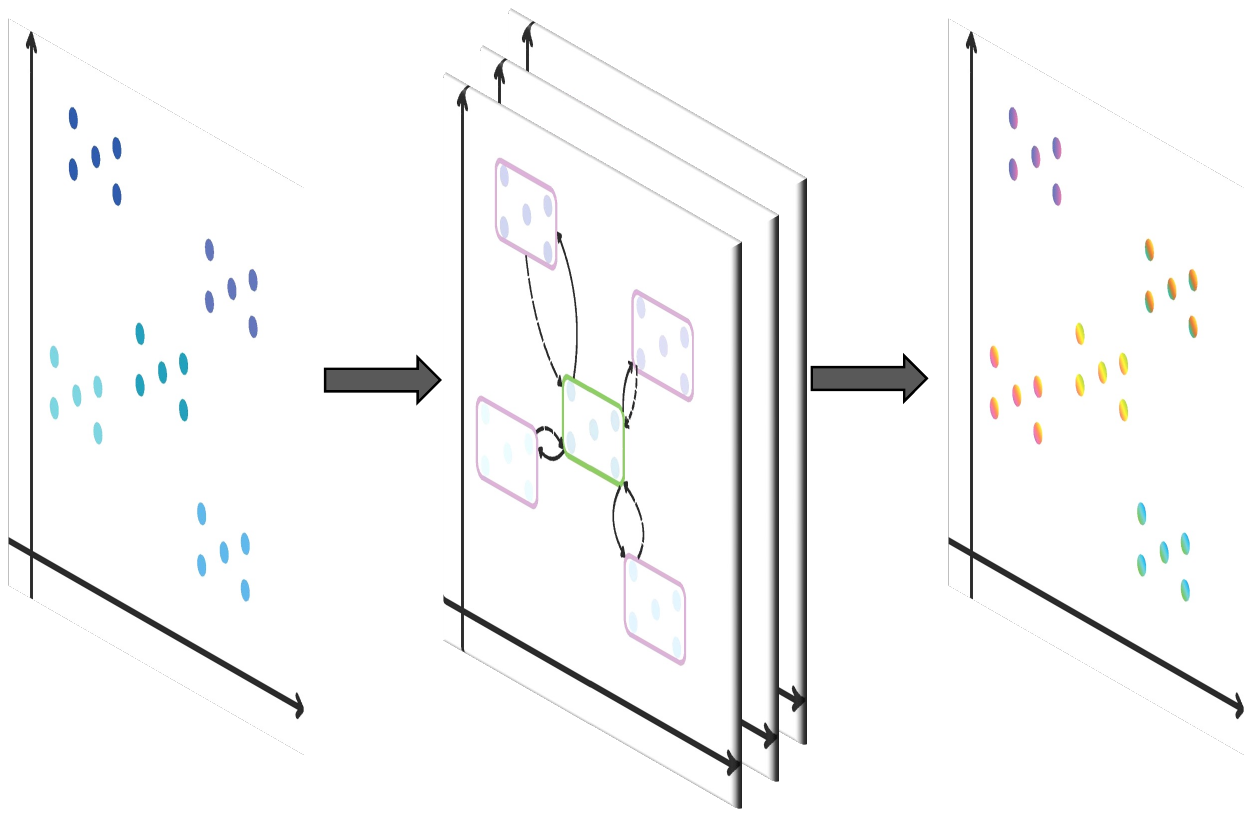}}
    \caption{ARMATA nodes encoder.}
    \label{fig:nodes_encoder}
\end{figure*}

The nodes encoder maps the $f$-dimensional feature matrix $X$ to a $d$-dimensional latent representation $X_{1} \in \mathbb{R}^{n \times d}$, as illustrated in \figurename~\ref{fig:nodes_encoder}, capturing an embedding of the input graph, providing a better characterization of the areas' formation. 

To obtain this graph representation, the nodes encoder uses a Gated Graph Convolutional Network~\cite{GConvNET}. This GNN, comprising $L$ layers, uses anisotropic aggregation and gating mechanisms to refine node and edge features through recursive message passing, improving graph representation. Hence, the node feature $h_{i}^\ell$ and edge feature $e_{ij}^\ell$ at layer $\ell$ are
\begin{eqnarray}
\begin{aligned}
&h_i^{\ell+1} = h_i^{\ell} + \\
&\operatorname{ReLU}\Biggl(\operatorname{BN}\Biggl(U_n^{\ell} h_i^{\ell} + \operatorname{AG}_{j \in \mathcal{N}_i}\Biggl(\sigma\left(e_{i j}^{\ell}\right) \odot V_n^{\ell} h_j^{\ell}\Biggr)\Biggr)\Biggr),
\end{aligned}
\label{h_l}
\end{eqnarray}

\begin{eqnarray}
\begin{aligned}
e_{ij}^{\ell+1} = &e_{ij}^{\ell} + \\
&\operatorname{ReLU}\Biggl(\operatorname{BN}\Biggl(A_n^{\ell} e_{ij}^{\ell}+ B_n^{\ell} h_i^{\ell} +C_n^{\ell} h_j^{\ell}\Biggr)\Biggr),
\end{aligned}
\label{e_l}
\end{eqnarray}
\noindent where $U_n^\ell, V_n^\ell, A_n^\ell, B_n^\ell, C_n^\ell \in \mathbb{R}^{d \times d}$ denote the learnable parameters of the nodes encoder, $\operatorname{ReLU}$ is the Rectified Linear Unit function, $\operatorname{BN}$ denotes batch normalization layer, $\operatorname{AG}$ is the neighborhood aggregation operator, $\sigma$ is the sigmoid function, and $\odot$ is the Hadamard product. For the first layer ($\ell=0$), the inputs $h_i^0$ and $e_{ij}^0$ are $d$-dimensional linear projections of the feature matrix $X$ and the Euclidean distance between the centers of the areas $\| p_i -p_j\|$, respectively. 

\subsubsection{Agents encoder}\label{subsubsec:agent_enc}
\begin{figure*}
    \centering
    \resizebox{\textwidth}{0.4\textwidth}{\includegraphics{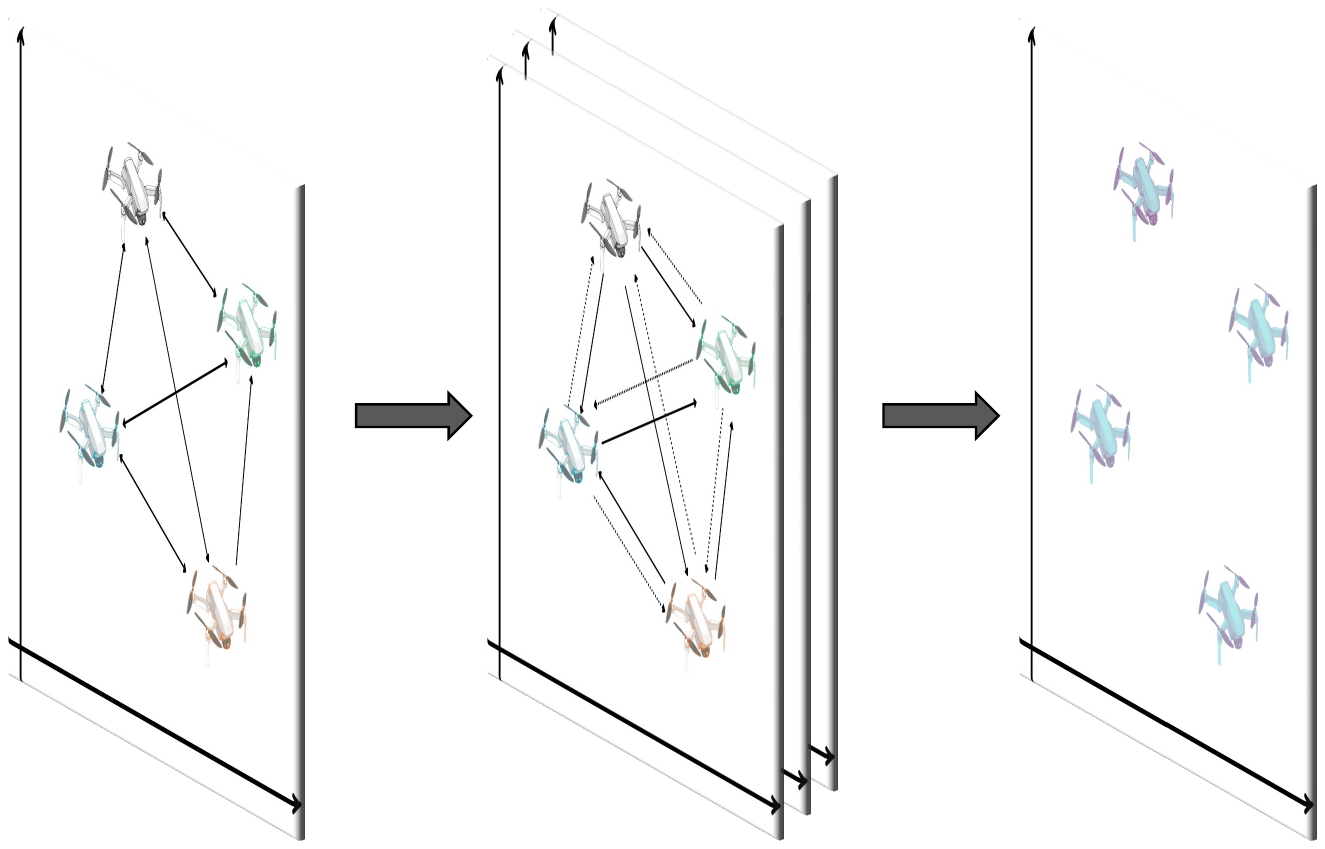}}
    \caption{ARMATA agents encoder.}
    \label{fig:agents_encoder}
\end{figure*}

The agents encoder is responsible for mapping the $2$-dimensional position matrix $Z$ to a $d$-dimensional latent representation $Z_{1} \in \mathbb{R}^{m \times d}$, as illustrated in \figurename~\ref{fig:agents_encoder}, capturing an embedding of the agents' locations, providing a better characterization of the agents' relative formation. 

The agents encoder has the same structure of the nodes encoder. It uses a GNN, comprising $L$ layers, to refine the agents' positions through recursive message passing, improving agents' relative representation. Hence, the agent embedding $h_{\alpha}^\ell$ and agent edge feature $e_{\alpha\beta}^\ell$ at layer $\ell$ are
\begin{eqnarray}
\begin{aligned}
&h_\alpha^{\ell+1} = h_\alpha^{\ell} + \\
&\operatorname{ReLU}\Biggl(\operatorname{BN}\Biggl(U_m^{\ell} h_i^{\ell} + \operatorname{AG}_{\beta \in \mathcal{M}_\alpha}\Biggl(\sigma\left(e_{\alpha \beta}^{\ell}\right) \odot V_m^{\ell} h_\beta^{\ell}\Biggr)\Biggr)\Biggr),
\end{aligned}
\label{agent_h_l}
\end{eqnarray}

\begin{eqnarray}
\begin{aligned}
e_{\alpha \beta}^{\ell+1} = &e_{\alpha \beta}^{\ell} + \\
&\operatorname{ReLU}\Biggl(\operatorname{BN}\Biggl(A_m^{\ell} e_{\alpha \beta}^{\ell}+ B_m^{\ell} h_\alpha^{\ell} +C_m^{\ell} h_\beta^{\ell}\Biggr)\Biggr),
\end{aligned}
\label{agent_e_l}
\end{eqnarray}
\noindent where $U_m^\ell, V_m^\ell, A_m^\ell, B_m^\ell, C_m^\ell \in \mathbb{R}^{d \times d}$ denote the learnable parameters of the agents encoder.




\subsection{Decoder}\label{sec:decoder}
\begin{figure*}
    \centering
    \resizebox{\textwidth}{0.4\textwidth}{\includegraphics{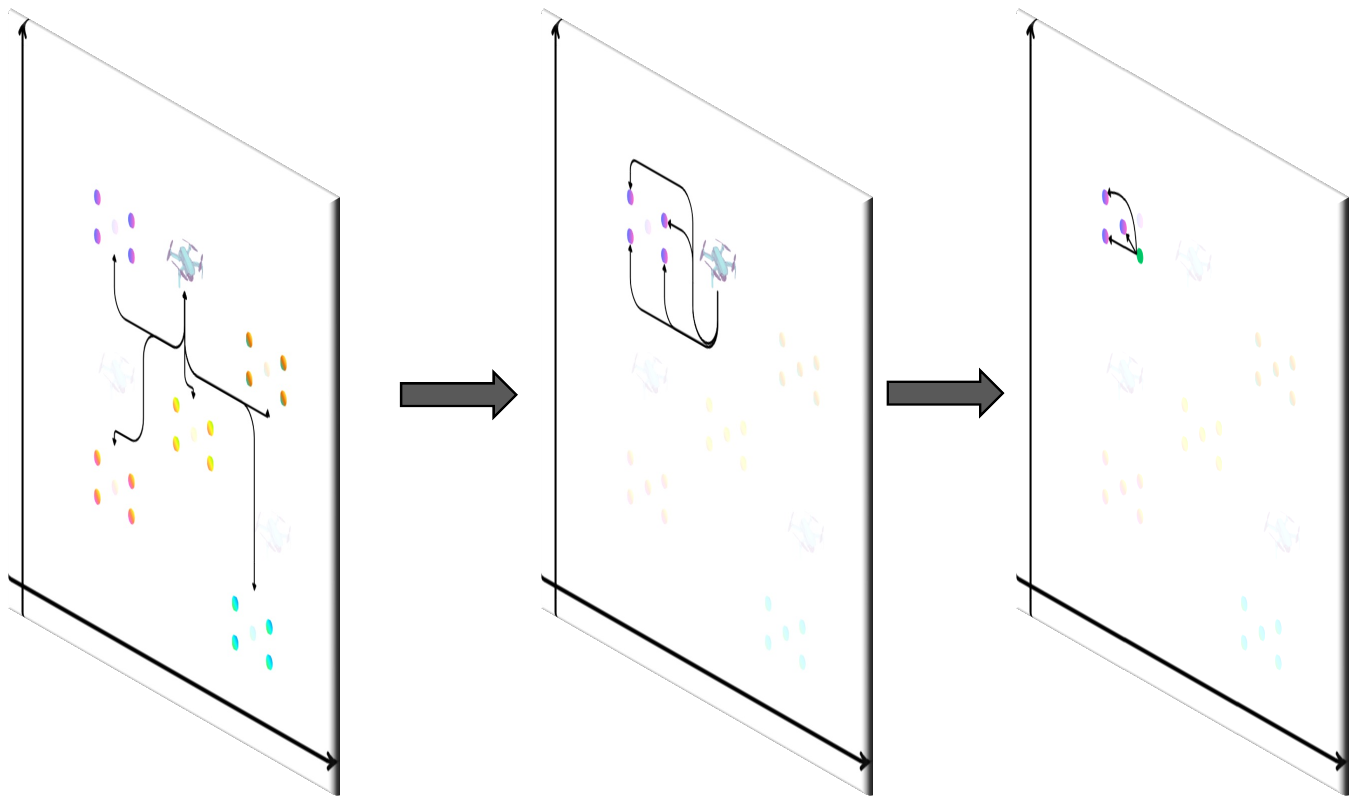}}
    \caption{ARMATA decoder.}
    \label{fig:decoder}
\end{figure*}

\begin{figure}
    \begin{center}
        \includegraphics[width=\columnwidth]{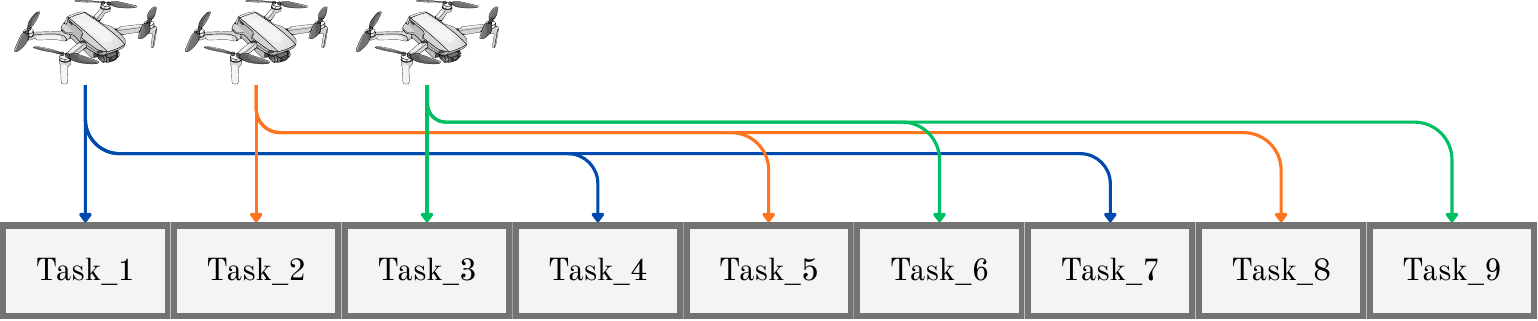}
    \end{center}
    \caption{ARMATA working principle.}
    \label{fig:Dec_functionality}
\end{figure}

As discussed earlier, task assignment is performed sequentially: agents are fed to the decoder one at a time in a fixed order, and each agent gets a new assignment. An assignment specifies the area to be scanned, the starting point, and the scanning pattern, as illustrated in \figurename~\ref{fig:decoder}. Analogous to large language models (LLMs), the decoder learns an optimal assignment sequence conditioned on previously generated decisions, as illustrated in \figurename~\ref{fig:Dec_functionality}. Using novel context representations, the model simultaneously learns which areas each agent should visit and the order in which they should be serviced.

Thereby, the decoder consists of three sequential Transformer networks: 
\begin{enumerate}
    \item \textbf{Network I} assigns an area to the input agent.
    \item \textbf{Network II} determines the starting point in the area.
    \item \textbf{Network III} decides the movement pattern for the area.
\end{enumerate}


Each network follows the same processing pipeline: input embedding, multi-head attention (MHA), embedding, attention, and softmax.

Network [I] uses the output of both components of the encoder, $h_{i}^L$ and $h_{\alpha}^L$, to build an initial context representation $\hat{h}_{(\alpha,i)}^C$ for the current area $i$ where agent $\alpha$ is located at assignment time step $t$. This context, given by
\begin{eqnarray}
\begin{aligned}
    &\hat{h}_{(\alpha,i)}^C = W_{F} \left[ h_{G}, h_{\pi_{1}^{\prime}}^L, h_{\pi_{t-1}^{\prime}}^L, h_{\alpha}^L, h_{\alpha}^{i,L}, h_i^L \right], \\
    \quad &h_{G} = \frac{1}{n} \sum_{i=1}^{n} h_{i}^L,
\end{aligned}
\label{eq:h_c_init}
\end{eqnarray}
\noindent includes learnable weights $W_{F} \in \mathbb{R}^{d \times d}$ and averages the features from the encoder. Here, $h_{\pi_{1}^{\prime}}^L$ and $h_{\pi_{t-1}^{\prime}}^L$ denote the features of the first and last areas in the partial tour ($\pi_\pi^{\prime}$), which contains the sequence of areas assigned up to time $t-1$. $h_{\alpha}^L$ is the encoding of the starting location of agent $\alpha$, $h_{\alpha}^{i,L}$ is the encoding of agent's $\alpha$ current location, and $h_{i}^L$ is the encoded feature vector of area $i$ which represents the current area of agent $\alpha$. The resulting context vector $\hat{h}_{(\alpha,i)}^C$ highlights the correlation among the current area of agent $\alpha$, the starting location of agent $\alpha$, the current location of agent $\alpha$, previously assigned areas, and the overall graph structure.

The initial context \eqref{eq:h_c_init} is refined using MHA over the node embeddings. We define the parameters of the first MHA ($\texttt{MHA}_1$) as $Q_1 = W_{C_1} \hat{h}_{(\alpha,i)}^C$, $K_1 = \left\{W_{A_1}[h_{1}^L, \ldots, h_{n}^L]\right\}$, and $V_1 = \left\{W_{A_2}[h_{1}^L, \ldots, h_{n}^L]\right\}$, where $W_{C_1}, W_{A_1}, W_{A_2} \in \mathbb{R}^{d  \times d}$ are learnable parameters. Here, $Q_1$ is a linear transformation of the input information, while $K_1$ and $V_1$ represent the available options as an output. The refined context is obtained as
\begin{equation}
h_{(\alpha,i)}^C=\operatorname{MHA}_{1}(Q_1, K_1, V_1).
\label{eq:MHA_1}
\end{equation}

The unnormalized logit for assigning area $j$ (or edge $e_{ij}$) to agent $\alpha$ is computed using an attention mechanism between the context $h_{(\alpha,i)}^C$ from \eqref{eq:MHA_1} and the embedding $h_{j}^L$:
\begin{eqnarray}
\hat{p}[A^\alpha_j|A^\alpha_i]= M \cdot \tanh \left(\frac{\left(W_{Q_1} h_{(\alpha,i)}^C\right)^T \cdot\left(W_{K_1} h_{j}^L\right)}{\sqrt{d_2}}\right),
\label{eq:P_j}
\end{eqnarray}
\noindent where $j \neq \pi_{t}^{\prime}$. If area $j$ has already been assigned or previously visited, $\hat{p}[A_j|A_i] = \infty$. $W_{Q_1}, W_{K_1} \in \mathbb{R}^{d \times d}$ are learnable parameters, and $\tanh$ keeps the logits in the range $[-M, M]$. The logits are then converted to probabilities $p[A^\alpha_j|A^\alpha_i]$ via a softmax operation.

Using \eqref{eq:P_j}, an area $j$ is now assigned to agent $\alpha$. Next, a starting point within area $j$ is selected by choosing one of its corners. This decision is handled by Network [II], which learns the correlation between the features of area $j$ and the location of agent $\alpha$ to determine an appropriate starting point. 

The ($x,y$) coordinates of the selected area's corner features and the current location of agent $\alpha$ are supplied to Network [II], which builds a context representation for the agent's location over area's $j$ features using MHA. This context captures the correlation between the agent's location and the candidate starting locations within the area. The agent's current location and the area's corner coordinates are embedded using linear feedforward layeers, resulting in $d_\Lambda$-dimensional embeddings:
\[
h_{st} = W_{C_2}h_{current}, \quad W_{\Lambda}[h_1,h_2,h_3,h_4]
\]
\noindent where \( h_{current} = (x_{current}, y_{current}) \), \( [h_1, h_2, h_3, h_4] \) are the corner coordinates of area $j$, and \( W_{C_2}, W_{\Lambda} \in \mathbb{R}^{d_\Lambda \times 2} \) are learnable parameters. $\texttt{MHA}_2$ parameters are:
\begin{eqnarray*}
    Q_2 &=& h_{st},\\
    \quad K_2 &=& \left\{W_{\Lambda 1}[h_1,h_2,h_3,h_4]\right\}, \\
    \quad V_2 &=& \left\{W_{\Lambda 2}[h_1,h_2,h_3,h_4]\right\}.
\end{eqnarray*}

The MHA operation yields the context \( h_{st}^C \):
\begin{equation}
    h_{st}^C=\operatorname{MHA}_{2}(Q_2, K_2, V_2).
\label{eq: h_s_C}
\end{equation}

The unnormalized logit for choosing corner \( h_k \), \( k \in \{1,2,3,4\} \), as the starting point is calculated using an attention mechanism between \( h_{st}^C \) and \( h_k \):
\begin{equation}
\begin{aligned}
&\hat{p}[\Lambda_k|A^\alpha_j]]=
&M \cdot \tanh \left(\frac{\left(W_{Q_2} h_{st}^C\right)^T \cdot\left(W_{K_2} h_k\right)}{\sqrt{d_\Lambda}}\right).
\end{aligned}
\label{eq: P_k}
\end{equation}

Finally, the logits \( \hat{p}[\Lambda_k|A^\alpha_j] \) are converted to probabilities \( p[\Lambda_k|A^\alpha_j] \) using a softmax operation.

The final decision is to select the scanning pattern that agent $\alpha$ should follow. This step mirrors previous stages. The coordinates of the candidate stopping points for each pattern ($h_{P_z}$) along with the selected starting point $h_k$ from Network [II], are provided to Network [III], which learns their correlation to select and appropriate pattern.

Following the same principles as the previous networks, Network [III] constructs a context representation for the starting point $h_k$ using MHA over the candidate stopping points associated with the patterns. Assuming $p$ possible patterns, the coordinates of $h_k$ and the stopping points of patterns $P_1, P_2, \ldots, P_p$ are embedded into a $d_\Psi$-dimensional latent space. $\texttt{MHA}_3$ parameters are:
\begin{eqnarray*}
    Q_3 &=& W_{C_3}[h_k], \\
    \quad K_3 &=& \left\{W_{\Psi_1}[h_{P_1},h_{P_2},\cdots,h_{P_p}]\right\},\\
    \quad V_3 &=& \left\{W_{\Psi_2}[h_{P_1},h_{P_2},\cdots,h_{P_p}]\right\}
\end{eqnarray*}
\noindent where \(W_{C_3}, W_{\Psi_1}, W_{\Psi_2} \in \mathbb{R}^{d_\Psi \times 2}\) are learnable parameters. The resulting context representation is \begin{equation}
    h_k^C=\operatorname{MHA}_3(Q_3, K_3, V_3).
    \label{eq:hK}
\end{equation}

The unnormalized logit for selecting pattern $P_z$, $z \in \{1,\dots,p\}$, is then computed using an attention operation between the context $h_k^C$ and the corresponding stopping point $h_{P_z}$:
\begin{equation}
\begin{aligned}
&\hat{p}[\Psi_z| A^\alpha_j,\Lambda_k]=
&M \cdot \tanh \left(\frac{\left(W_{Q_3} h_k^C\right)^T \cdot\left(W_{K_3} h_{P_z}\right)}{\sqrt{d_\Psi}}\right),
\end{aligned}
\label{eq:P_p}
\end{equation}
\noindent where $W_{Q_3} \in \mathbb{R}^{d_\Psi \times d_\Psi}$ and $W_{K_3} \in \mathbb{R}^{d_\Psi \times 2}$ are learnable parameters. The logits $\hat{p}[\Psi_z| A^\alpha_j, \Lambda_k]$ are normalized with softmax over all patterns. The probability of assigning area $A_j$, starting point $h_k$, and pattern $P_z$) to agent $\alpha$ is thus
\begin{equation}
p[j,k,z|\alpha]= p[A^\alpha_j] \cdot p[\Lambda_k|A^\alpha_j] \cdot p[\Psi_z| A^\alpha_j, \Lambda_k].
\label{eq:P_a}
\end{equation}

This procedure is repeated until all $n$ areas are assigned and scanned. This process consists of $n/m$ assignment round, each comprising $m$ sub-steps in which each agent is assigned exactly to one new task. At the end of every sub-step, the new agents' locations are encoded, through the agents encoder, to be used for the next assignment round. After $n$ total steps, all areas and their corresponding tasks are distributed among all agents.
 

\subsection{RL Policy Training}
RL is highly effective for training networks in scenarios where ground truth solutions are unavailable, as in the case of search and rescue multi-agent task assignment. In this paper, we use the REINFORCE policy gradient algorithm due to its simplicity. The objective is to minimize the total distance covered by all the agents in a given scenario, which in turn would minimize the total execution time of the mission. 

The loss for an instance $s$, which represents an assignment scenario, parameterized by $\theta$ is defined as $\mathcal{L}(\theta|s)=\mathbb{E}_{p_{\theta}(\pi|s)}[L(\pi)]$, where $L(\pi)$ is the total tour length of all the agents and $p_{\theta}(\pi|s)$ is the assignment probability distribution of the tour and agents $\pi$.

The gradient for minimizing $\mathcal{L}$ using REINFORCE is:
\begin{equation}
    \nabla \mathcal{L}(\theta|s) = \mathbb{E}_{p_{\theta}(\pi|s)}[(L(\pi)-b(s))\nabla \operatorname{log}(p_{\theta}(\pi|s))],
    \label{gradient}
\end{equation}
where $b(s)$ is a baseline to reduce gradient variance. Here, $L(\pi)$ is the total tour length per epoch, and $p_{\theta}(\pi|s)$ is the probability of the tour and its assignment, calculated as the product of action probabilities $p[j,k,z|\alpha]$ from \eqref{eq:P_a}.

While REINFORCE serves as our primary algorithm due to its effectiveness, we also investigate Proximal Policy Optimization (PPO) as a comparative policy gradient method. PPO is widely utilized in reinforcement learning literature for its ability to constrain policy updates, offering a trade-off between training stability and complexity~\cite{schulman2017ppo}. Motivated by its proven effectiveness in related domains, we test the performance quality when using PPO for training our policy.

In a typical PPO framework, a surrogate objective function is maximized rather than directly minimizing a loss based on a single sample. The surrogate objective function includes a probability ratio and an advantage estimate. The probability function is defined as:
\begin{equation}
    r_\theta(\pi|s) = \frac{\pi_\theta}{\pi_{\theta_{old}}}
    \label{eq:prob_ratio_ppo}
\end{equation}
with $\pi_\theta$ being the current policy and $\pi_{\theta_{old}}$ being the policy from the previous iteration. This ratio is used to measure the divergence between the new policy and the old one and to prevent large policy divergences. The other part of the surrogate function is the advantage estimate and it is defined as:
\begin{equation}
    \hat{A} = b(s) - L(\pi)
    \label{eq:advantage}
\end{equation}
with $b(s)$ being a baseline function and $L(\pi)$ is the total tour length of all the agents. In typical PPO implementations, the advantage will be defined as $\hat{A} = L(\pi) - b(s)$ as the algorithm will try to maximize this advantage. Nonetheless, we look to minimize the tour loss, so the advantage is defined as in~\eqref{eq:advantage}. Hence, if the advantage is negative, then the policy obtained a better solution than the baseline, and vice versa.

The resulting PPO objective will then be:
\begin{equation}
Obj_{\texttt{PPO}}(\theta) = \mathbb{E}{p_{\theta_{\text{old}}}(\pi|s)} \left[ \min \left( r \hat{A}, \ \operatorname{clip}(r, 1-\epsilon, 1+\epsilon) \hat{A} \right) \right],
\label{eq:obj_PPO}
\end{equation}
where $\epsilon$ is a hyperparameter controlling the clipping range. As mentioned earlier, the clipping is used here to prevent large policy deviations.
\section{Simulation Setup}\label{sec:SimulationSetup}

\subsection{Dataset}
For training, \( n \) areas, represented by frames, as well as \(m\) agents were generated within a unit square. Each agent was generated at a random ($x,y$) distinct position. Similarly, each area center was sampled from a uniform distribution $C = (c_x,c_y)$ where  $c_x, c_y \in [0,1]$. Then a random radius $r \in [0.01, 0.03]$ was added and subtracted from the center coordinates to form the four corners. This process was repeated until all $n$ areas were generated, ensuring they stayed within the unit square and did not overlap neither with each other nor with the agents. The unit square was chosen for easy scalability. In total, the dataset comprises $128,000$ randomly generated maps per epoch. 

\subsection{Hyper-parameters}\label{subsec:parameters}
The encoder consists of three layers with a hidden dimension of $d=128$. The decoder comprises the three components discussed in Section~ \ref{sec:decoder}. For the area assignment probability $p[A^\alpha_j|A^\alpha_i]$ in \eqref{eq:P_j}, we used a hidden dimension of $d=128$ and $8$ attention heads. For the starting-point and pattern-selection probabilities in \eqref{eq: P_k} and \eqref{eq:P_p}, we set $d_\Lambda=d_\Psi=128$ and use a single attention head in each respective MHA module. Overall, this configuration resulted in approximately $360,000$ trainable parameters.  

\subsection{Training Process}
The model was trained for $100$ epochs, with each epoch comprising $128,000$ randomly generated samples, processed in mini-batches. Each sample represents a distinct multi-agent task assignment scenario, and all samples within a batch share the same number of agents and the same number of areas. The number of agents per sample was uniformly sampled from [$5,10$], while the number of areas ranged from [$35,100$]. To simplify evaluation, the number of areas in each instance was constrained to be an integer multiple of the number of agents, ensuring an equal number of tasks per agent.

The RL environment was developed using OpenAI Gym \cite{GYM}, chosen for its compatibility with the Robot Operating System (ROS) \cite{2017_gym_gazebo_ros, 2019_gym_gazebo}, facilitating future deployment on real robotic platforms.
The model was implemented in PyTorch 2.1 and trained on an NVIDIA GeForce RTX-3090 GPI with an Intel i9 11900KF eight-core CPU (3.5GHz base, up to 5.3GHz with Turbo Boost).

\subsection{Evaluation}\label{subsec:evaluation}
To evaluate the proposed framework, we compared its assignments against solutions obtained from Google OR-Tools~\cite{ortools}. OR-Tools is a widely used open-source combinatorial optimization suite that supports linear, mixed-integer, and constraint programming, with specialized solvers for routing and scheduling. Its scalability, accessibility, and broad adoption in both academic and industrial settings make it a suitable baseline for comparison. 

For evaluation, we generated four test datasets, each containing $1280$ samples. Two datasets had samples similar to those encountered during the training and are referred to as \textit{in-range}. One of these datasets included scenarios with $5$ agents and $40$ areas, while the other had samples with $8$ agents and $72$ areas. The remaining two were \textit{out-of-range} and were designed to assess  generalization to unseen problem scales, with configurations of $4$ agents and $24$ areas, and $12$ agents and $120$ areas, respectively.   
\section{Results and Discussion}\label{sec:results}
This section presents a comprehensive evaluation of the proposed centralized autoregressive framework for solving the joint multi-agent task assignment problem described in Section~\ref{subsec:evaluation}. 
\subsection{Fixed vs Varying Starting Locations}
In this test, the starting positions of the agents were fixed across the batch samples for one run, while they were varied for the other run. The hyperparameters are shown in \tablename~\ref{tab:hyperparameters} under the column ``\textbf{Varying locations}''.

\begin{figure*}
    \centering
    \begin{subfigure}[b]{\columnwidth}
        \centering
        \includegraphics[width=\columnwidth]{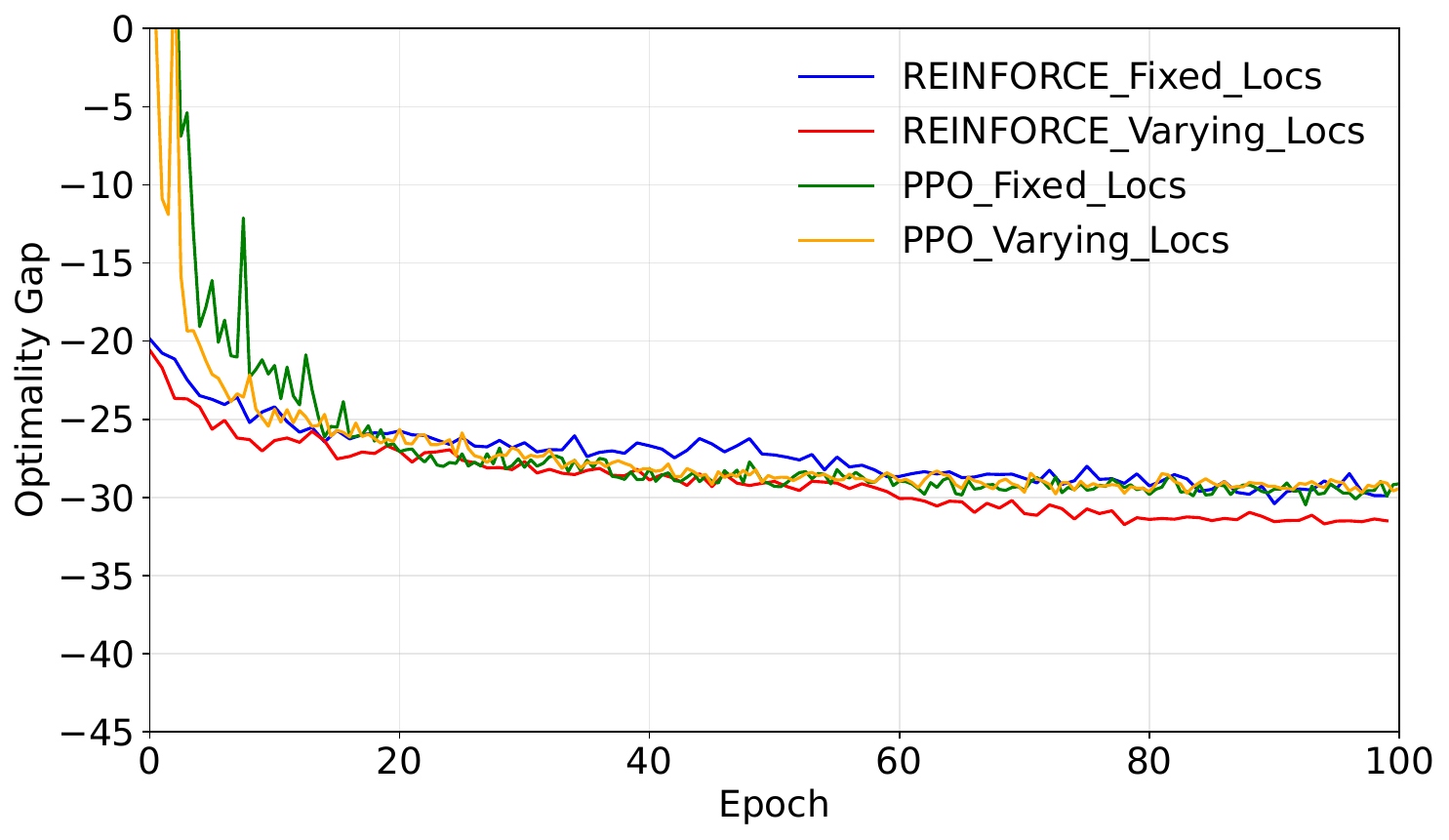}
        \caption{Optimality gap for in range dataset with 40 areas and 5 agents.}
        \label{fig:A_in1}
    \end{subfigure}
    \begin{subfigure}[b]{\columnwidth}
        \centering
        \includegraphics[width=\columnwidth]{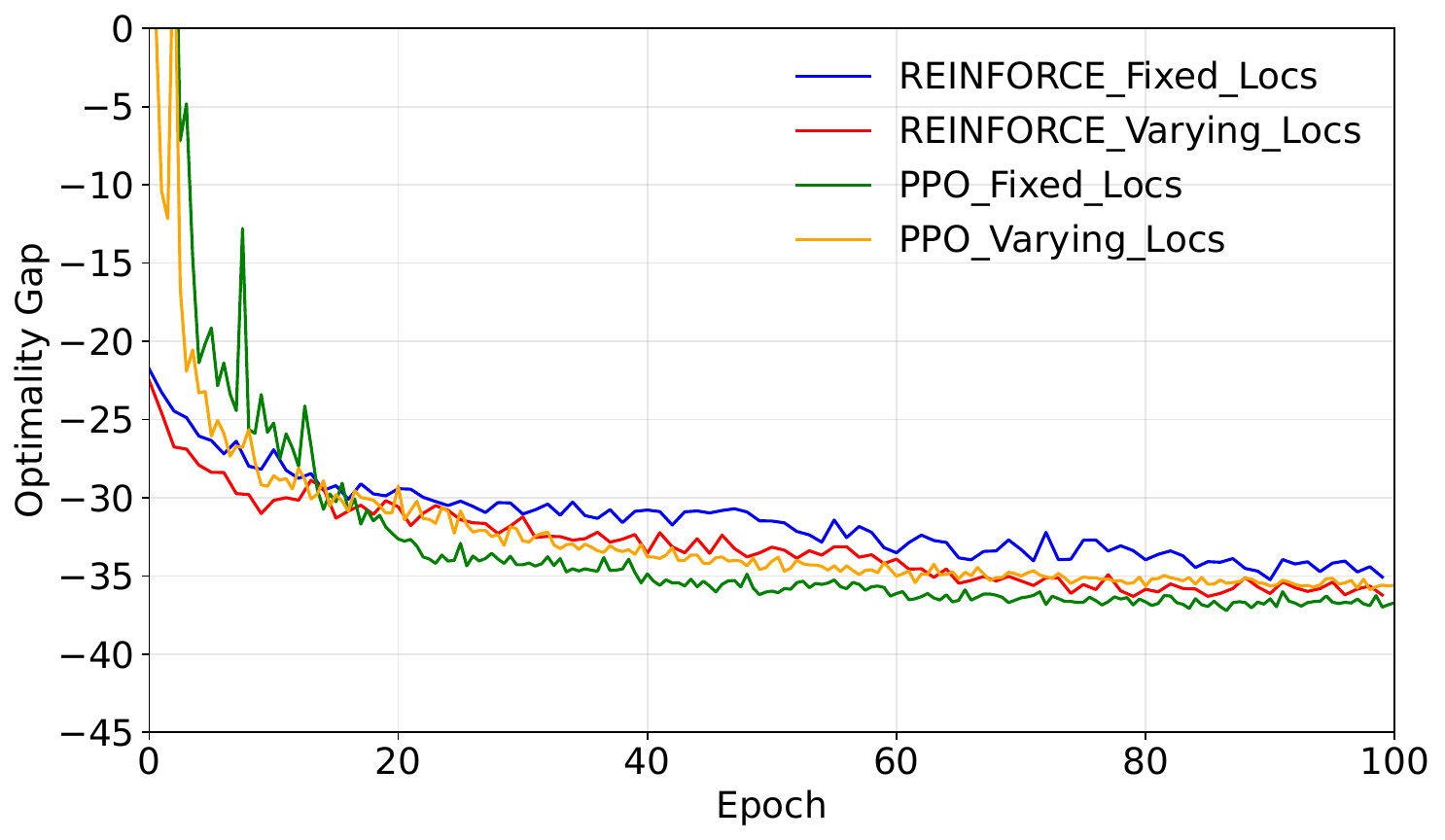}
        \caption{Optimality gap for in range dataset with 72 areas and 8 agents.}
        \label{fig:A_in2}
    \end{subfigure}
    \begin{subfigure}[b]{\columnwidth}
        \centering
        \includegraphics[width=\columnwidth]{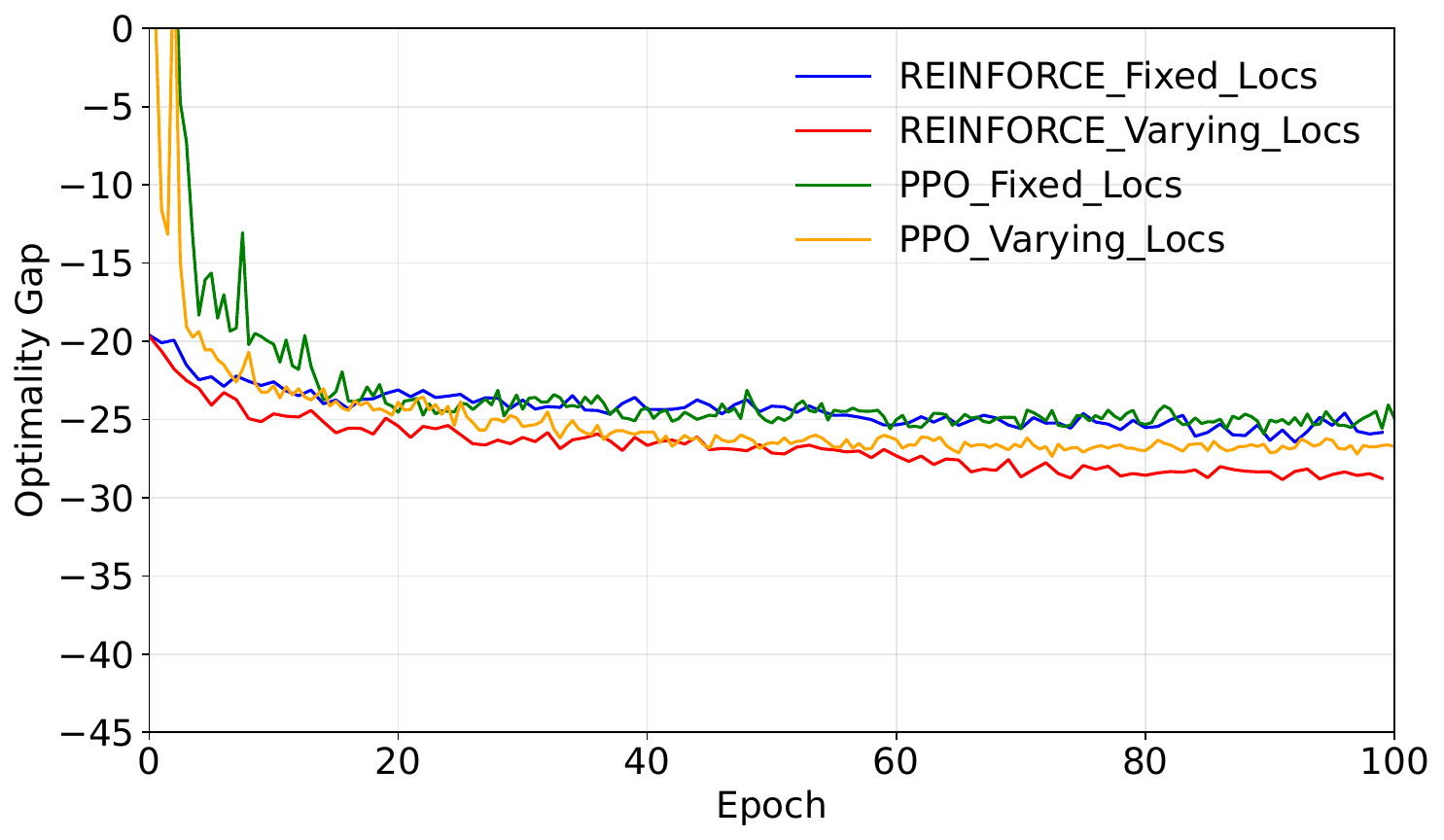}
        \caption{Optimality gap for out range dataset with 24 areas and 4 agents.}
        \label{fig:A_out1}
    \end{subfigure}
    \begin{subfigure}[b]{\columnwidth}
       \centering
       \includegraphics[width=\columnwidth]{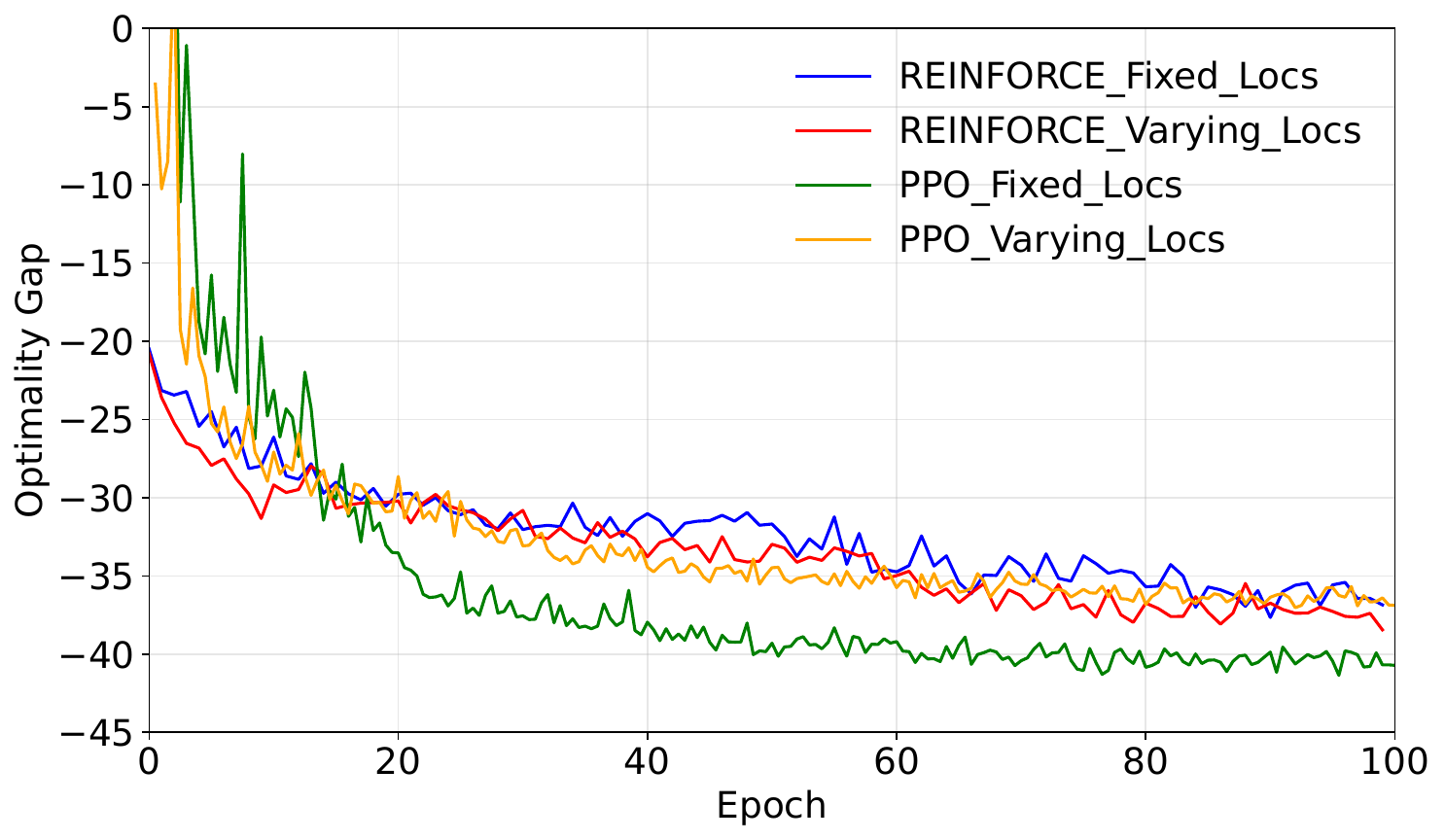}
       \caption{Optimality gap for out range dataset with 180 areas and 12 agents.}
        \label{fig:B_out2}
    \end{subfigure}

    \caption{Performance of ARMATA when the agents' starting locations are varied vs when they are fixed.}
    \label{fig:fixedSL_vs_varyingSL}
\end{figure*}

As shown in \figurename~\ref{fig:fixedSL_vs_varyingSL}, both approaches resulted in very similar performance. Varying the starting locations consistently resulted in better results with REINFORCE, while it was case-dependent with PPO. For the smaller datasets, varying the starting locations yielded better performance. On the other hand, fixing the starting locations resulted in better results with the larger datasets.

\subsection{Disabling the Agents Encoder}\label{subsec:results_II}
In this test, the agents encoder was disabled and the starting positions of the agents were fixed across the batch samples for one run, while they were varied for the other run. The hyperparameters are shown in \tablename~\ref{tab:hyperparameters} under the column ``\textbf{No agents encoder}''. 

\begin{table*}[t]
\centering
\caption{Training and model hyperparameters.}
\label{tab:hyperparameters}
\begin{tabular}{@{}llll@{}}
\toprule
\textbf{Hyperparameter} & \textbf{Varying locations} & \textbf{No agents encoder} 
\\ \midrule
Batch size & $128$ & $128$\\
Normalization type & Batch & Batch\\
Number of epochs & $100$ & $100$\\
Number of encoder layers & $3$ & $3$\\
Number of attention heads & $8$ & $8$\\
Learning rate & $10^{-4}$ & $10^{-4}$\\
Decaying rate & -- & -- \\ \bottomrule
\end{tabular}
\end{table*}

\begin{figure*}
    \centering
    \begin{subfigure}[b]{\columnwidth}
        \centering
        \includegraphics[width=\columnwidth]{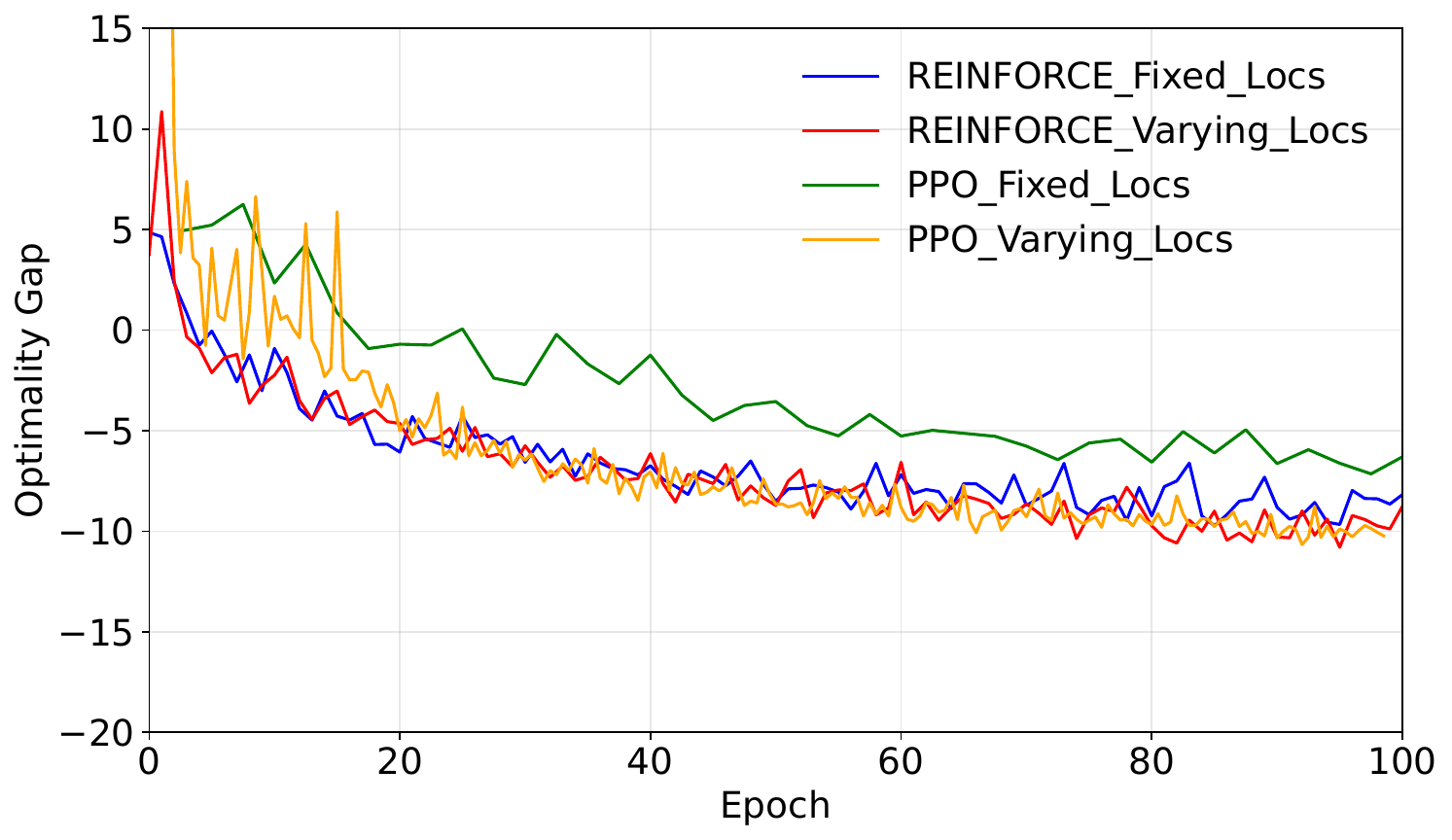}
        \caption{Optimality gap for in range dataset with 40 areas and 5 agents.}
        \label{fig:B_in1}
    \end{subfigure}
    \begin{subfigure}[b]{\columnwidth}
        \centering
        \includegraphics[width=\columnwidth]{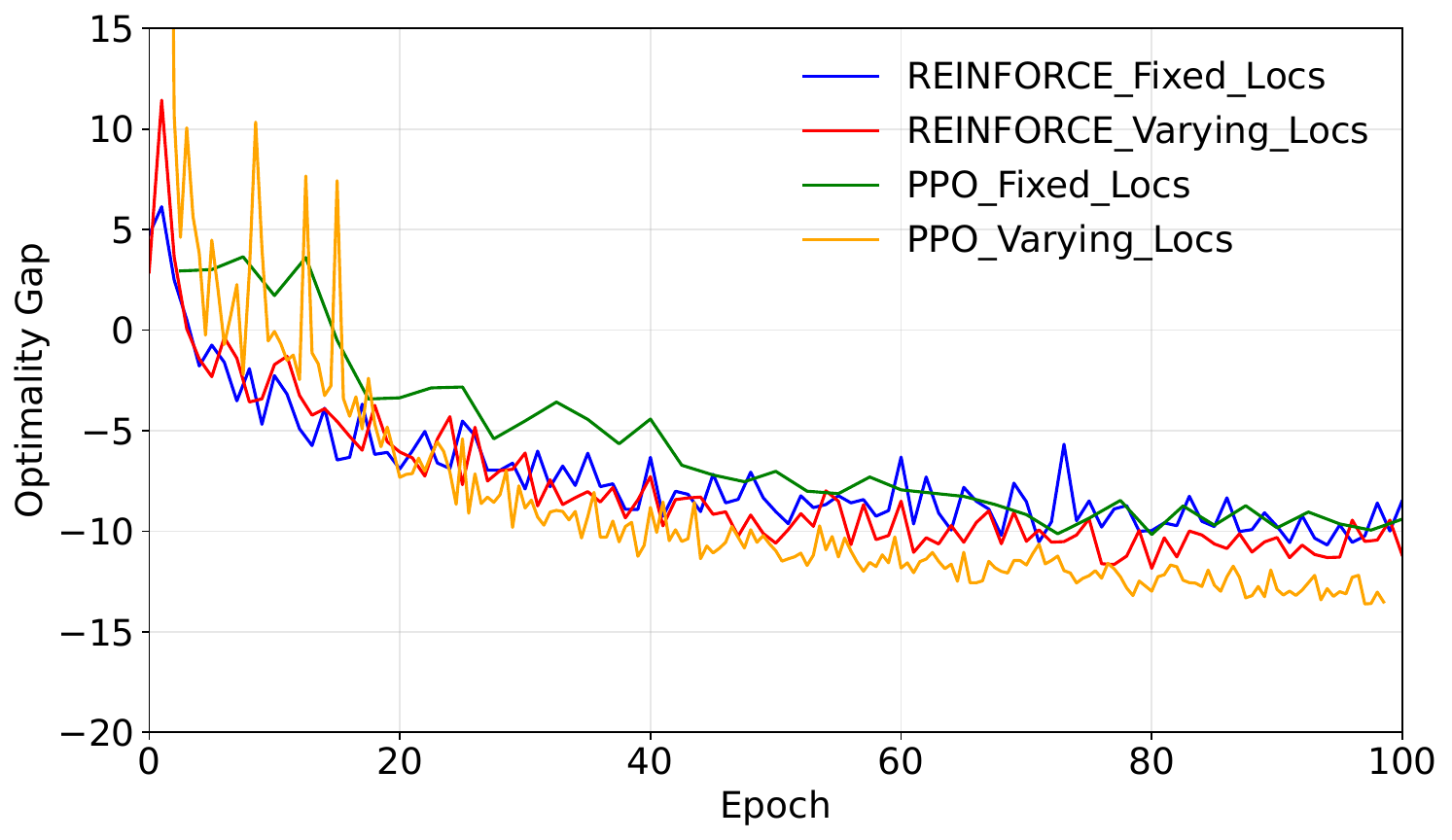}
        \caption{Optimality gap for in range dataset with 72 areas and 8 agents.}
        \label{fig:B_in2}
    \end{subfigure}
    \begin{subfigure}[b]{\columnwidth}
        \centering
        \includegraphics[width=\columnwidth]{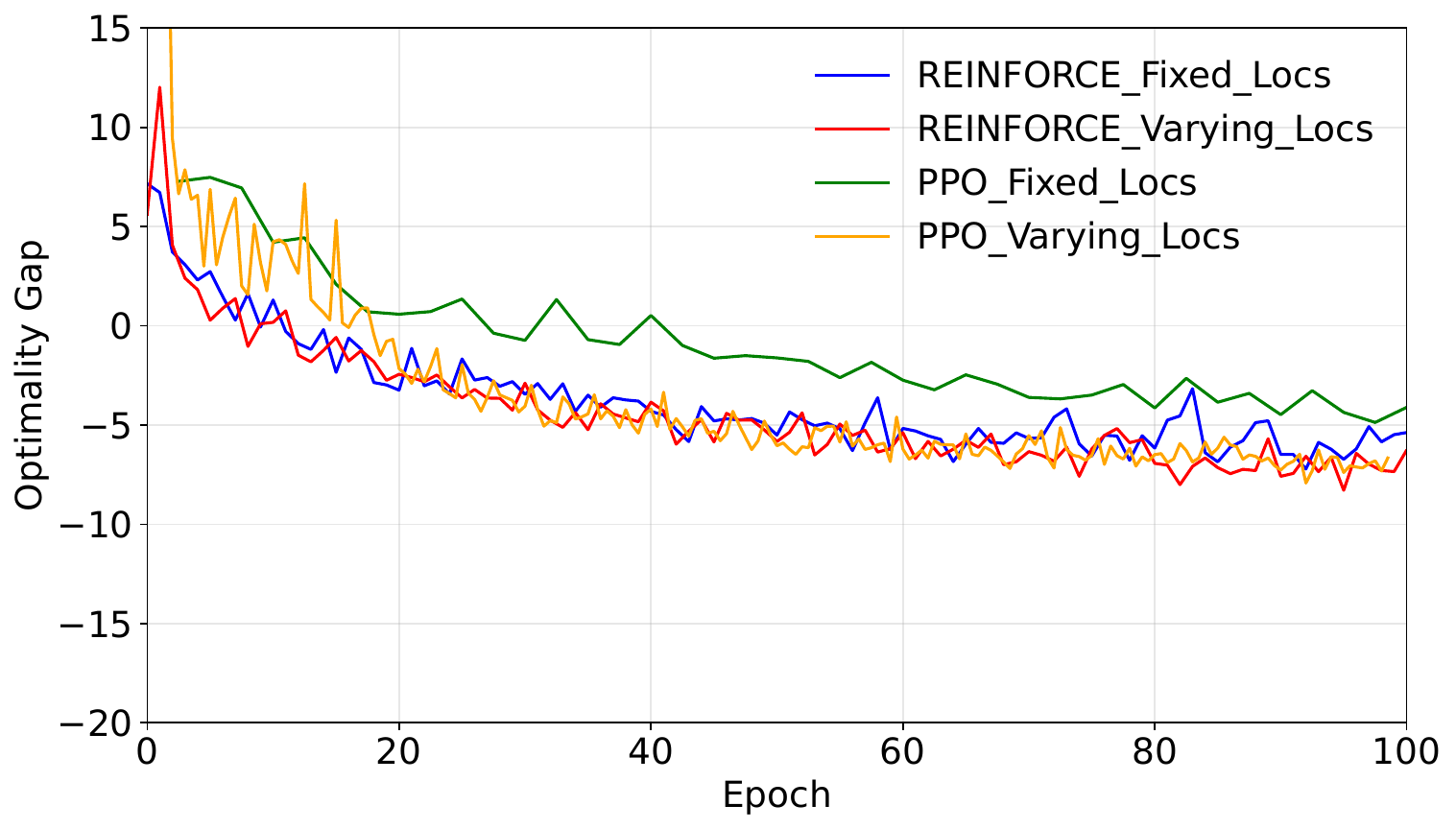}
        \caption{Optimality gap for out range dataset with 24 areas and 4 agents.}
        \label{fig:B_out1}
    \end{subfigure}
    \begin{subfigure}[b]{\columnwidth}
       \centering
       \includegraphics[width=\columnwidth]{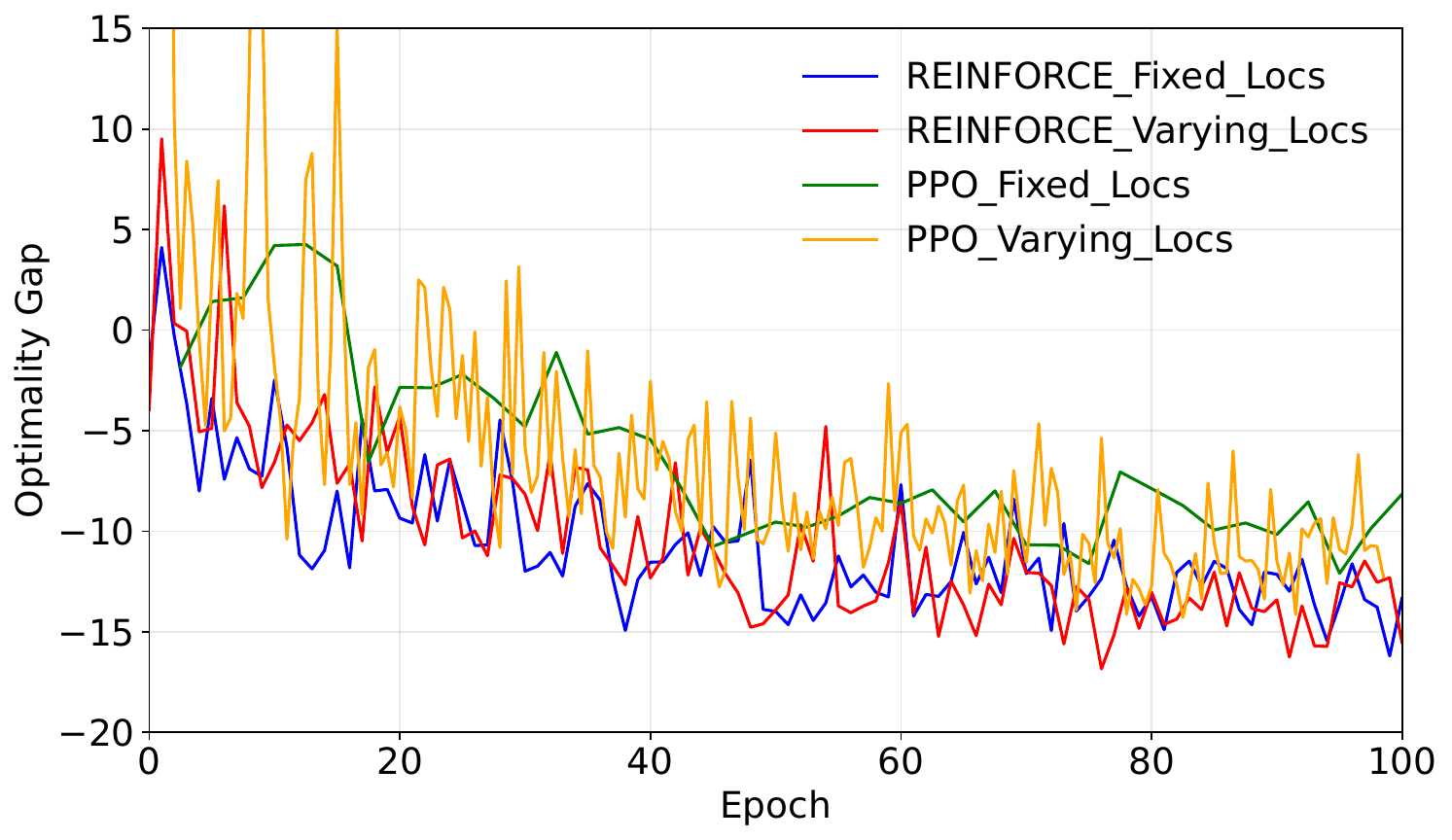}
       \caption{Optimality gap for out range dataset with 180 areas and 12 agents.}
        \label{fig:B_out2}
    \end{subfigure}

    \caption{Performance of ARMATA when the agents encoder is disabled.}
    \label{fig:No_agent_encoder}
\end{figure*}




As shown in \figurename~\ref{fig:No_agent_encoder}, the framework's performance degraded significantly when the agents encoder was removed, as the optimality gap increased by approximately 20\%. These results are expected since the agents encoder enables the decoder to better comprehend the problem since the agents encoder provides infromation about the agents' formation relative to each other.
\subsection{Comparison with LKH3 and CPLEX}\label{subsec:CPLEX}
To further assess the proposed framework, we compared its performance with an industrial-grade optimization tool, IBM CPLEX~\cite{IBM_CPLEX}, as well as with the state-of-the-art metaheuristic Lin-Kernighan-Helsgaun-3 (LKH-3)~\cite{Helsgaun_2017_LKH3}. LKH3 is widely regarded as a leading metaheuristic for routing problems, while IBM CPLEX provides high quality solutions for multi-agent task allocation and path planning, as reported in~\cite{2025_end_to_end_learning}. However, the datasets considered in this work are substantially larger and involve more agents than those studied in~\cite{2025_end_to_end_learning}, requiring additional modifications to enable CPLEX to produce feasible solutions.

For instances with $24$ areas and $4$ agents, CPLEX was able to find a solution when provided with the full problem states. Nonetheless, for larger instances, such as those with $40$ areas and $8$ agents, supplying the fully connected graph resulted in memory overload and system failure. To overcome this issue, a pruning strategy was used in which only the $k$-nearest neighbors were kept for each node of the graph, rather than using a fully connected graph. This allowed CPLEX to obtain feasible solutions for the $40$-area, $8$-agent dataset. For even larger instances, specifically those with $72$ and $180$ areas and $8$ and $12$ agents, respectively, the pruning strategy did not help. Hence, an alternative initialization-based approach was used. Random feasible solutions were first generated, and the best among them was chosen as an initial solution for CPLEX. CPLEX was then initialized with this solution and allowed to attempt further optimization. Empirically, however, CPLEX was unable to improve upon the provided initial solutions for these large-scale instances. LKH-3 was able to find the solution for all the test cases when provided with the full problem without any required modifications.

The results in \tablename~\ref{tab:cplex_results} show that while all three approaches generate feasible tours, ARMATA achieves superior performance both in terms of solution quality and computational efficiency, particularly for large-scale instances. ARMATA yielded tours that are $14\%-26 \%$ better than the solutions obtained from the other solvers while improving the computational time by up to $3$ orders of magnitude. 

\begin{table*}
\centering
\caption{Average tour lengths abtained from CPLEX, LKH3, and ARMATA.}
\label{tab:cplex_results}
\begin{tabular}{@{}cccccccc@{}}
\toprule
\multirow{2}{*}{\textbf{\begin{tabular}[c]{@{}c@{}}Number of \\ Areas\end{tabular}}} & \multirow{2}{*}{\textbf{\begin{tabular}[c]{@{}c@{}}Number of \\ Agents\end{tabular}}}& \multicolumn{2}{c}{\textbf{LKH3}}& \multicolumn{2}{c}{\textbf{CPLEX}} & \multicolumn{2}{c}{\textbf{ARMATA}} \\ \cmidrule(l){3-8} 
& & $\mu$ [m] & $\mu$\_time [s] 
& $\mu$ [m] & $\mu$\_time [s] 
& $\mu$ [m] & $\mu$\_time [s] \\ \midrule
24 & 4 &  5.9028 & 1.159 & 8.3154 & 123.75 & \textbf{5.0628} & \textbf{0.0029} \\
40 & 5 & 7.4732 & 2.276 & 11.0650 & 101.25 & \textbf{6.5128} & \textbf{0.0061} \\
72 & 8 & 11.5958 & 46.097 & 12.7275 & 123.75 & \textbf{9.1108} & \textbf{0.0166} \\
180 & 12 & 24.3931 & 152.20 & 20.3547 & 405 & \textbf{14.8783} & \textbf{0.0872} \\ \bottomrule
\end{tabular}
\end{table*}

A key advantage of ARMATA lies in its generalization capability, which is particularly significant given that recent end-to-end frameworks, such as the approach in~\cite{2025_end_to_end_learning}, often exhibit limited robustness outside their training distributions and struggle to scale to unseen problem instances.

An even more pronounced strength of ARMATA is its execution efficiency. As problem size increases, the  computational cost of both LKH3 and CPLEX grows substantially. For the largest instance with $180$ areas, CPLEX required over $400~s$ and LKH3 more than $150~s$ to converge, whereas ARMATA computed the solution in only $0.0165~s$. Across all evaluated scenarios, ARMATA consistently operated in the sub-second range, achieving speedups of several orders of magnitude over the benchmark methods. This substantial reduction in computation time highlights ARMATA’s suitability for real-time dynamic applications, demonstrating that high scalability and rapid re-planning can be achieved without a significant loss in solution quality.
\section{Conclusion}\label{sec:conclusion}
This paper introduced the first centralized, fully end-to-end autoregressive framework for jointly addressing task allocation and routing in multi-agent systems. By means of a novel multi-stage decoding mechanism, the proposed architecture unifies high-level workload distribution and low-level trajectory planning within a single decision process. Extensive empirical evaluation demonstrated that the framework consistently outperformed heuristic solvers, achieving up to a $35\%$ improvement in solution quality over well established  solvers such as Google OR-Tools, IBM CPLEX and LKH-3, while reducing computation time by several orders of magnitude. Moreover, in contrast to recent end-to-end learning approaches, the proposed model exhibits strong generalization to unseen problem instances without retraining. Overall, by combining high solution quality with sub-second execution times, the framework offers a scalable and effective solution for real-time multi-agent coordination in complex, spatially distributed environments.

\begingroup
\bibliographystyle{IEEEtran} 
\bibliography{references}
\endgroup

\end{document}